# Lattice Boltzmann simulations of the pool boiling curves above horizontal heaters with homogenous and heterogeneous wettability

*Wandong Zhao, Jianhan Liang\*, Mingbo Sun\*, Xiaodong Cai*

Science and Technology on Scramjet Laboratory, National University of Defense Technology, Changsha 410073, China

**\*Corresponding authors**
*Email: jhleon@vip.sina.com (Jianhan Liang)*
*Email: sunmingbo@nudt.edu.cn (Mingbo Sun)*

**Highlights:**

The boiling curves on homogenous and heterogeneous surfaces were investigated.

Increasing the wettability of hydrophilic wall results in a high critical high flux.

Effects of hydrophilic-hydrophobic mixed wettability on boiling curves were studied.

Decreasing the wettability of mixed wall yields an upward shift of the boiling curve.

Decreasing the width of mixed wall results in a leftward shift of the boiling curve.

**Abstract:**

A hybrid thermal lattice Boltzmann phase-change model was performed to simulate the pool boiling above a smooth horizontal heater. The effects of homogenous and heterogeneous wettability on entire boiling curves from the onset of nucleate boiling (ONB) to fully developed film boiling were investigated comprehensively. Results show that concerning the homogeneous and hydrophilic heater, increasing the wall wettability results in a high critical heat flux (CHF) and yields a rightward and upward shift of the boiling curve. It is also found that decreasing the wettability promotes the occurrence of ONB at a lower overheat, but it also makes the boiling process move into a film boiling regime with a lower degree of wall superheat. Regarding the homogeneous and hydrophobic heater, there are quite lower CHF and shorter transition boiling regime, whereas the CHF point and film boiling occur at a lower wall superheat. For the heterogeneous and hydrophilic heater, increasing the difference of wettability yields a leftward shift of the boiling curve. Furthermore, the effects of surface wettability and the width between two hydrophilic-hydrophobic mixed surface on boiling curves were studied, which indicated that decreasing the wettability of the hydrophobic surface produces an upward shift of boiling curve and decreasing the width generates a leftward and upward shift of the nucleate boiling pattern. Moreover, the enhancement of boiling heat transfer in terms of hydrophilic-hydrophobic mixed surfaces was also quantitively verified.

**Keywords:**

Phase-change lattice Boltzmann method; Boiling heat transfer; Boiling curve; Mixed wettability; homogenous and heterogeneous

# 1. Introduction

    As one of the highest efficient means to remove a large quantity of thermal capacity, boiling heat transfer (BHT) has been extensively capitalized on daily life and industrial applications [1, 2]. In the meanwhile, the boiling curve of the BHT is considered as the most useful characteristics to understand the boiling process after the first well-known experiments on pooling boiling conducted by the pioneer worker Nukiyama [3]. Numerous efforts have been made to augment the BHT enhancement and understand the mechanisms of the BHT. Thereby, Many experiments have been conducted to perform the different effects on the boiling curves, such as the properties of heating plate or working fluid [4, 5], the structure of heating surface [6, 7] and the heating conditions [8-10]. Among these, the characteristics of heating surface are regarded as the most prime influences on the BHT [11, 12], especially for the onset of nucleate boiling (ONB) and critical heat flux (CHF). Aa a consequence, a large number of researchers have focused on the surface characteristics on the BHT including the effects of heating surface wettability, roughness or structures [12, 13].

    The surface characteristic of heater is viewed as a dominate factor to enhance the BHT due to the improvement of the nucleation site density, the nucleate bubble growth and the departure frequency of bubbles [12]. In particular, the surface wettability of heating plate can be tailored to induce significantly different BHT performances including bubble nucleation, ONB and CHF [13-17]. The former investigates have reported that an early ONB would form in the hydrophobic surface with a lower wall superheat [16, 18], whereas it also leads to a lower CHF due to the stronger relationship between the generated bubble and the hydrophobic surface [19]. On the other hand, the hydrophilic surface with an excellent wettability results in a higher CHF compared to hydrophobic surface, while it also produces drawback that the occurrence of BHT would require a high wall superheat. Therefore, in recent years, a large number of researchers have highlighted in the BHT enhancement by introducing a heterogeneous heating surface with hydrophobic-hydrophilic mixed or mixed wettability. [11, 18, 20-23].

    Betz et al. [20] presented an experimental study using microlithography approaches to manufacture a smooth and horizontal heterogeneous surface composed of hydrophilic and hydrophobic networks for pool BHT. They revealed that the heterogeneous surface has an extremely high CHF and heat transfer efficiency compered to homogenous and hydrophilic heating surface. And the hydrophilic networks of the heterogeneous surface can block the occurrence of an insulating vapor film. Bourdon et al. [18] designed a "T"-shaped polished bronze heating plate to experimentally investigate the effects of surface roughness and wettability on ONB. They reported that the surface wetting variation at nanoscale can promote the BHT and decrease the requirement of wall superheat, and the BHT only take place in the hydrophobic grafted regions. Almost at the same time, a significant investigation was conducted by Jo et al. [11], who experimentally quantify that a heterogeneous heating plate consisted of hydrophobic spots on a hydrophilic surface leads to a better nucleate BHT than the homogenous wetting surface. Subsequently, Dai et al. [21] conducted an experimental study on pool BHT to understand the different nucleate BHT processes occurred in hydrophobic-hydrophilic heaters with a wide range of wettability, which is fabricated from functionalized multiwall carbon nanotubes by coupled hydrophilic spots on the pristine multiwall carbon nanotube. It was confirmed that the new type of interface could be superior in improving nucleate BHT owing to effective nucleation site density, bubble growth and departure frequency formed by this kind of surface.

Jo and co-authors recently performed several experimental studies on BHT with a hydrophilic substrate mixed with hydrophobic dots. The effects of spatially-difference characteristics on nucleate BHT and CHF were investigated by adjusting the diameter of hydrophobic dots and the relative length between two spots. They observed that the diameter of hydrophilic spots, the relative length between two spots and the quantity of hydrophilic spot are the prime influence factors on BHT, whereas the ratio of region blanketed by hydrophobic pattern interfaces has no effect on the BHT [22]. In the meanwhile, they also carried out an experimental research of the BHT on a heterogeneous wettability microstructure with a self-assembled monolayers (SAMs) and micro-post structure. The result showed that the growth of nucleation site is induced by the micro-post regimes, and bubble growth is affected by SAMs and micro-post structures, leading to an enhancement of the BHT [23]. Kumar et al. [24] experimentally studied the BHT to gain insights on the influence of the wettability of heterogeneous surfaces considering three types of wettability. In their work, the bubble dynamic, local heat flux and heat transfer coefficient were comprehensively studied based on the heterogeneous cylindrical heating surface. They confirmed that there is a leftward shift in the pool boiling curve for the heterogeneous surface owing to the high nucleation site density and less departure bubble time.

Most recently, numerical simulation has also exhibited a powerful potential to cope with the pool BHT, so that scholars also presented numerical simulations on the effect of surface wettability on BHT to improve the BHT performance. Li et al [25]. conducted a numerical simulation of the pool BHT occurred in a type of textured surface with hydrophilic-hydrophobic mixed wettability. They confirmed the mechanism of the BHT enhancement of the mixed wettability surface compared to the homogeneous wetting surface, and the effects of pillar height and width on heat flux and heat transfer coefficient were also discussed. They numerically obtained a leftward shift of the boiling curve and a leftward and upshift of the heat transfer coefficient curve from the BHT occurred in the homogeneous, heterogeneous and hydrophilic-hydrophobic mixed wettability surfaces. Furthermore, Ma et al. [26] recently performed numerical simulations of the pool BHT for the four types of micro-pillar heaters. In their simulations, the effects of structure and surface wettability on the BHT were studied. It was found the mixed wettability heater having hydrophobic pillar in the hydrophilic substrate presents a best BHT performance than other three types of micro-pillar heaters.

To date, although the above-mentioned investigations have demonstrated the enhancement performance of the BHT by studying the ONB and CHF, these literature only focused on the part of characteristics of the BHT rather than studied the entire boiling curves including the transition boiling and film boiling regime. Thereby, to better explore the characteristics of the BHT occurred in both homogenous and heterogeneous superheat heater, the current work are dedicated to study the entire boiling curve from the ONB to fully developed film boiling on the homogenous and heterogeneous heating surfaces with different wettability. Additionally, the effects of different wettability and structures of the heterogeneous surface on the whole boiling curves are also discussed in detail. The lattice Boltzmann method (LBM), which has been considered as a powerful discrete solver based on the gas kinematic dynamics, is employed in current investigation. This mean has also been developed as a powerful tool to capture and cope with the various and complicated multiphase flows [27, 28]. The hybrid thermal pseudopotential LBM with an ability to simulate the two-phase change proposed by Li et al. [29] is used in the current work. This model has been extensively utilized to numerically simulate the pool boiling heat transfer above a smooth and horizontal superheat surface, and the different boiling patterns (ONB, CHF and fully developed film boiling) have been successfully obtained [29]. In the meanwhile, the dynamics characteristics of Leidenfrost droplet [30] and pinning-depinning mechanism of droplet's evaporation [31, 32] above a superheat surface were also successfully investigated by employing the hybrid phase-change LBM. Therefore, this model employed in current simulation has reasonable accuracy and reliability to handle the BHT.

## 2. The hybrid pseudopotential phase-change lattice Boltzmann model

*2.1. The improved pseudopotential multiple-relaxation-time lattice Boltzmann model for liquid-vapor flows*

The pseudopotential LBM model developed by Shan-Chen [33, 34] is widely utilized in multiphase flows. However, in this model, the evolution of density distribution function (DF) with SRT operator [35] was introduced, which had some drawbacks in numerical stability and accuracy [36]. Recently, Li et al. [37] improved the source term in the moment space, and the flow evolution of density DF with MRT operator has a form as [25, 38-41]

$$f_\alpha(\boldsymbol{x}+\boldsymbol{e}_\alpha\delta_t, t+\delta_t) = f_\alpha(\boldsymbol{x},t) - \overline{\Lambda}_{\alpha\beta}(f_\beta - f_\beta^{eq})\big|_{(x,t)} + \delta_t F'_\alpha\big|_{(x,t)} \tag{1}$$

where $f$ and $f^{eq}$ denote the density DF and its equilibrium DF, respectively, and the quantities $\delta_x$ and $\delta_t$ are the lattice space and the time space [29]. $e_a$ indicates the discrete velocity and $F'_\alpha$ is the forcing term. In current simulation, the D2Q9 model is employed [42]. Meanwhile, $\overline{\Lambda} = \boldsymbol{M}^{-1}\Lambda\boldsymbol{M}$ in Eq. (1) is the collision matrix. $\boldsymbol{M}$ is the orthogonal transfer matrix, and $\Lambda$ is the diagonal relaxation matrix, giving [37, 43]

$$\begin{aligned}\Lambda &= diag(s_0, s_1, s_2, s_3, s_4, s_5, s_6, s_7, s_8) \\ &= diag(\tau_\rho^{-1}, \tau_e^{-1}, \tau_\varsigma^{-1}, \tau_j^{-1}, \tau_q^{-1}, \tau_j^{-1}, \tau_q^{-1}, \tau_\upsilon^{-1}, \tau_\upsilon^{-1})\end{aligned} \tag{2}$$

where $s_1 = s_2$, $s_3 = s_5$, $s_7 = s_8$. The flow non-dimensional relaxation time associated with kinematic viscosity ($\upsilon$) can be calculated by [44]

$$\tau_\upsilon = \frac{1}{s_7} = \upsilon/c_s^2 + 0.5 \tag{3}$$

Furthermore, the relaxation time has a relationship as [44, 45]:

$$\tau_v = \tau_V + \frac{\rho-\rho_V}{\rho_L-\rho_V}(\tau_L - \tau_V) \tag{4}$$

where the quantities $V$ and $L$ represent the gas and liquid phase, respectively.

By employing transformation matrix $\boldsymbol{M}$, the DF $f$ and $f^{eq}$ can be projected into the moment space $\boldsymbol{m} = \boldsymbol{M}\cdot f$, $\boldsymbol{m}^{eq} = \boldsymbol{M}\cdot f^{eq}$, thus $\boldsymbol{m}^{(eq)} = \rho\left[1, -2+3(u_x^2+u_y^2), \rho-3(u_x^2+u_y^2), u_x, -u_x, u_y, -u_y, u_x^2-u_y^2, u_xu_y\right]$ [46, 47], where $u_x, u_y$ is the component of velocity $\boldsymbol{u} = \sqrt{u_x^2+u_y^2}$, therefore Eqs. (5) and (6) can be obtained from Eq. (1) [48].

$$\boldsymbol{m}^* = \boldsymbol{m} - \Lambda(\boldsymbol{m}-\boldsymbol{m}^{eq}) + \delta_t(\boldsymbol{I}-\frac{\Lambda}{2})\overline{\boldsymbol{S}} \tag{5}$$

$$f_i(\mathbf{x}+\mathbf{e}_i\delta_t, t+\delta_t) = f_i^*(\mathbf{x},t) \tag{6}$$

Herein, $\bar{S} = MS$ is the external forcing term, $S = (S_0, S_1, S_2, S_3, S_4, S_5, S_6, S_7, S_8)^T$ and $f^* = M^{-1}m^*$. Thereby, the evolution of density DF can be solved by means of Eqs. (5) and (6), which are also called as "collision process" and "streaming process" in previous literature.

To realize the thermodynamic consistency and mechanical stability, Li et al. [37] introduced an external forcing term, giving

$$\bar{S} = \begin{bmatrix} 0 \\ 6(u_x F_x + u_y F_y) + \dfrac{12\varpi |\mathbf{F}_m|^2}{\psi^2 \delta_t (\tau_e - 0.5)} \\ -6(u_x F_x + u_y F_y) + \dfrac{12\varpi |\mathbf{F}_m|^2}{\psi^2 \delta_t (\tau_\varsigma - 0.5)} \\ F_x \\ -F_x \\ F_y \\ -F_y \\ 2(u_x F_x - u_y F_y) \\ (u_x F_y - u_y F_x) \end{bmatrix} \tag{7}$$

where $\varpi$ is employed to adjust numerical stability, and $\mathbf{F}_m = (F_{mx}, F_{my})$ is the interaction force term for the two-phase segregation, while $\mathbf{F} = (F_x, F_y)$ is the total force. In the pseudopotential LBM, Shan-Chen [33] proposed the interaction force for liquid-vapor separation, defined

$$\mathbf{F}_m = -G\psi(\mathbf{x},t)\left[\sum_i w(|\mathbf{e}_i|^2)\psi(\mathbf{x}+\mathbf{e}_i,t)\mathbf{e}_i\right] \tag{8}$$

where $G$ is the interaction strength, and $w(|\mathbf{e}_a|^2)$ is the weight factor [43, 48]. And $\psi$ in the Eq. (8) is defined as [49]

$$\psi = \sqrt{\dfrac{2(P_{EOS} - \rho c_s^2)}{Gc^2}} \tag{9}$$

where $P_{EOS}$ is equation of state for the real gas, and in present simulation, the Peng-Robinson (P-R) equation of state is used, giving

$$P_{EOS} = \dfrac{\rho RT}{1-b\rho} - \dfrac{a\varphi(T)\rho^2}{1+2b\rho - b^2\rho^2} \tag{10}$$

where $\varphi(T) = [1+(0.37464+1.54226\omega-0.26992\omega^2)(1-\sqrt{T/T_c})]^2$, and $w = 0.344$, $a = 0.45724 R^2 T_c^2 / P_c$, $b = 0.0778 RT_c / P_c$. The quantities $T_c$, $P_c$ denote the critical temperature and pressure. The quantities $a$, $b$ and $R$ are taken as 2/49, 2/21 and 1, respectively [29], therefore the critical temperature $T_c$ is assumed as 0.1094 for the current simulation. With respect to the MRT model, the macroscopic density and velocity have a form as [37]

$$\rho = \sum_i f_i, \quad \rho\mathbf{v} = \sum_i \mathbf{e}_i f_i + \dfrac{\delta_t \mathbf{F}}{2} \tag{11}$$

Recently, Li et al. introduced a new interaction force associated with pseudopotential force, which can adjust a wide range of surface wettability, defined by Eq. (12), and for more detailed information of the implementation in the LBM can be found in Refs. [44, 50].

$$\boldsymbol{F}_{ads} = -G_w \psi(\boldsymbol{x},t) \left[ \sum_i w(|\boldsymbol{e}_i|^2) \psi(\rho_w) s(\boldsymbol{x}+\boldsymbol{e}_i) \boldsymbol{e}_i \right] \tag{12}$$

where $s(\boldsymbol{x}+\boldsymbol{e}_i)$ is a switch scheme. The gravitational force $\boldsymbol{F}_g$ is given below

$$\boldsymbol{F}_g(\boldsymbol{x}) = (\rho(\boldsymbol{x}) - \rho_v)\boldsymbol{g} \tag{13}$$

where $\boldsymbol{g} = (0, -g)$ and $\rho_v$ is average density of the entire fluid domain, which had been widely adopted in former literature in terms of the boiling heat transfer simulation [4, 51, 52], hence the total force in Eqs. (7) and (11) is $\boldsymbol{F} = \boldsymbol{F}_m + \boldsymbol{F}_{ads} + \boldsymbol{F}_g$.

*2.2. Energy equation for heat transfer*

The phase change LBM was first proposed by Zhang and Chen [53]. The governing equation of temperature field considering the diffusion interface was given as follows

$$\rho \frac{De}{Dt} = -p\nabla \cdot \boldsymbol{v} + \nabla \cdot (\lambda \nabla T) \tag{14}$$

where $e = C_V T$ is internal energy, $C_V$ is the specific heat capacity at the constant volume, and $\lambda$ is the thermal conductivity [54]. The entropy's local equilibrium energy equation without considering viscous dissipation is given by

$$\rho T \frac{ds}{dt} = \nabla \cdot (\lambda \nabla T) \tag{15}$$

And the general relation of entropy is given by

$$ds = \left(\frac{\partial s}{\partial T}\right)_v dT + \left(\frac{\partial s}{\partial v}\right)_T dv \tag{16}$$

According to the Maxwell relationship

$$\left(\frac{\partial s}{\partial v}\right)_T = \left(\frac{\partial p_{EOS}}{\partial T}\right)_v \tag{17}$$

and based on the chain relation and the definition of specific heat capacity, it can be derived that

$$\left(\frac{\partial s}{\partial T}\right)_v = \frac{(\partial u/\partial T)_v}{(\partial u/\partial s)_v} = \frac{Cv}{T} \tag{18}$$

With the help of Eqs. (17) and (18), Eq. (19) can be rewritten according to Eq. (16).

$$ds = \frac{Cv}{T} dT + \left(\frac{\partial p}{\partial t}\right)_v dv = \frac{Cv}{T} dT + \left(\frac{\partial p_{EOS}}{\partial t}\right)_v d\left(\frac{1}{\rho}\right) = \frac{Cv}{T} dT - \frac{1}{\rho^2}\left(\frac{\partial p_{EOS}}{\partial t}\right)_v d\rho \tag{19}$$

Further on, the Eq. (15) can be obtained as

$$\rho Cv \frac{dT}{dt} = \frac{T}{\rho}\left(\frac{\partial p_{EOS}}{\partial t}\right)_\rho \frac{d\rho}{dt} + \nabla \cdot (\lambda \nabla T) \tag{20}$$

By employing the material derivative $D(\bullet)/Dt = \partial_t(\bullet) + \mathbf{v} \cdot \nabla(\bullet)$, Eq. (20) can be further rewritten as

$$\frac{\partial T}{\partial t} + \mathbf{v} \cdot \nabla T = \frac{1}{\rho C v} \nabla \cdot (\lambda \nabla T) + \frac{T}{\rho^2 C v} \left( \frac{\partial p_{EOS}}{\partial t} \right)_\rho \frac{d\rho}{dt} \tag{21}$$

As a consequence, the target temperature equation associated with equation of state can be deriver form Eq. (21) under the relation of continuity equation, giving

$$\frac{\partial T}{\partial t} = -\mathbf{v} \cdot \nabla T + \frac{1}{\rho C v} \nabla \cdot (\lambda \nabla T) - \frac{T}{\rho C v} \left( \frac{\partial p_{EOS}}{\partial t} \right)_\rho \nabla \cdot \mathbf{v} \tag{22}$$

The right side of Eq. (22) is marked with $K(T)$, and by using fourth-order Runge-Kutta scheme [55], the time discretization of the governing equation of energy can be solved by

$$T^{t+\delta t} = T^t + \frac{\delta t}{6}(h_1 + 2h_2 + 2h_3 + h_4) \tag{23}$$

where $h_1$, $h_2$, $h_3$ and $h_4$ are calculated by Eq. (24) respectively.

$$h_1 = K(T^t), h_2 = K(T^t + \frac{\delta_t}{2} h_1), h_3 = K(T^t + \frac{\delta_t}{2} h_2), h_4 = K(T^t + \delta_t h_3) \tag{24}$$

For a quantity $\phi$ in LBM model, the spatial gradient and second-order Laplace are defined as [56, 57]

$$\partial_i \phi(\mathbf{x}) \approx \frac{1}{c_s^2 \delta_t} \left[ \sum_a w_\alpha \phi(\mathbf{x} + \mathbf{e}_a \delta_t) \mathbf{e}_a \right] \tag{25}$$

$$\nabla^2 \phi(\mathbf{x}) \approx \frac{2}{c_s^2 \delta_t^2} \left[ \sum_a w_\alpha (\phi(\mathbf{x} + \mathbf{e}_a \delta_t) - \phi(\mathbf{x})) \right] \tag{26}$$

Generally, the multiphase flows are determined by MRT-LBM, whereas the transport of temperature field is calculated by means of finite-difference method, which is coupled by the $P_{EOS}$ in the Eq. (10).

## 3. Results and discussion

*3.1. Computational setup*

The schematic of the computational domain for the BHT considering the conjugate heat transfer is illustrated in Fig. 1. As shown in this figure, a rectangle fluid domain is set to be $2\lambda_d \times 1.5\lambda_d$, while an additional grid size of $2\lambda_d \times 0.15\lambda_d$ is taken as solid domain for the conjugate heat transfer below the fluid domain. Note that the $\lambda_d$ is the Taylor most-dangerous wavelength for the two different densities fluid flow [4, 58], which is applied for nondimensionalization of the computational domain, giving:

$$\lambda_d = 2\pi \sqrt{\frac{3\sigma}{g(\rho_l - \rho_v)}} \tag{27}$$

which is also obtained as

$$\lambda_d = 2\sqrt{3} l_0 \tag{28}$$

where the quantity $l_0$ is the characteristic length, which has been intensively adopted in previous investigations [4, 52, 59]. it represents the ratio of surface tension and buoyancy force, determined by:

$$l_0 = \sqrt{\frac{\sigma}{g(\rho_l - \rho_v)}} \quad (29)$$

where $\sigma$ is the surface tension. Moreover, the characteristic time $t_0$ is introduced in current research, which is given by

$$t_0 = \sqrt{l_0/g} \quad (30)$$

As shown in Fig. 1, the computational domains of the liquid and vapor phase are chosen as $2\lambda_d \times 0.75\lambda_d$. Note that a length $\lambda_d$ has high resolution to handle the simulation of the pool BHT as demonstrated in Ref. [4], so that a length of $2\lambda_d$ for current simulation has a rather high resolution. In current simulation, the periodic BC is imposed in the y-axis for the flow and temperature, whereas the no-slip flow is used in the solid domain. Thereby, the halfway bounce-back scheme [60] is employed for the entire solid domain to deal with the multiphase flows. Besides, constant temperature $T_{sat}$ and $T_b$ are assumed at the topper and lower surface, respectively.

Before the thermal energy release, the stagnant fluid is saturated with pool under the condition that the initial temperature of liquid and vapor phase is set to be $T_{sat} = 0.86T_c$, which indicates that the density for the liquid and vapor phase are equal to $\rho_L = 6.5$ and $\rho_V = 0.38$, respectively [29]. According to Refs. [25, 29, 53, 61], the physical properties of the liquid and vapor phase are chosen as follows: special heat $C_{V,L} = C_{V,V} = 6.0$, kinematic viscosity $\upsilon_L = 0.1$, $\upsilon_V = 0.5/3$, whereas the thermal conductivity $\lambda = \rho C_V \chi$ for the computational domain is chosen as the proportion of the density $\rho$ with $C_V \chi = 0.033$ [29]. The density for solid domain is assumed as $\rho_S = 3\rho_L$, and the thermal conductivity is taken as $\lambda_S = 16.25$. Thereby, in present simulation, the dynamic ratio of liquid and vapor phase is $\mu_L/\mu_V \approx 10$, and the thermal conductivity ratio of the liquid and vapor phase is $\lambda_L/\lambda_V \approx 17$, whereas the thermal conductivity ratio of solid domain and liquid phase is $\lambda_S/\lambda_L \approx 75$. It should be noted that, unless specific description, all above-mentioned properties for current simulation are kept the same in the next simulations for the BHT. At the same time, the previous simulation presented by Gong and Cheng [4] confirmed that the thermal conductivity of the heater has no influence on the CHF when the thermal conductivity ratio of solid and liquid domain is set to be with $\lambda_S/\lambda_L \approx 30 \sim 150$. Thereby, the current conductivity ratio of $\lambda_S/\lambda_L \approx 75$ can exclude the influence of the thermal conductivity ratio on the CHF. The gravitational acceleration is taken as $g = (0, -0.00005)$. Thus, the Taylor most-dangerous wavelength is computed to be $\lambda_d \approx 200$, so that the computation domain for the BHT is confirmed as $L_x \times L_y = 400 \times 300$. Note that, in this simulation, all quantities are based on lattice unit with the lattice constant $c = \delta_x/\delta_t = 1$, $\delta_x = \delta_t = 1$. Moreover, following Ref. [62], the specific latent heat for the current fluid property is assumed as $h_{LV} = 0.58$ when the saturated temperature is set to be $T_{sat} = 0.86T_c$.

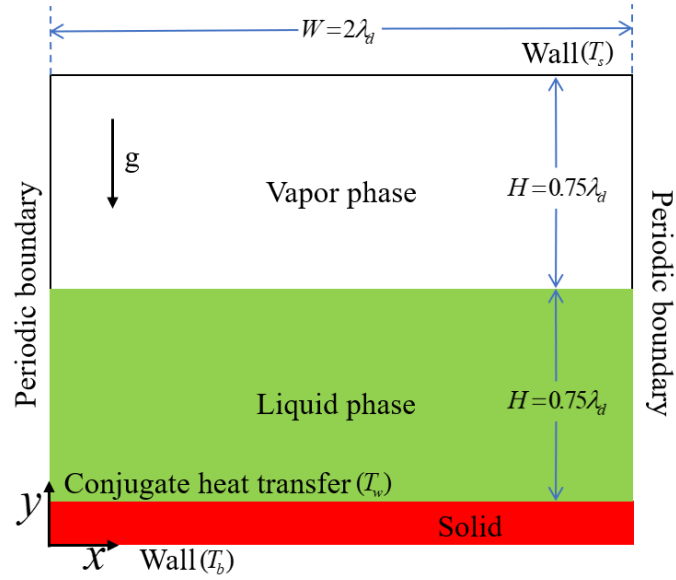

Fig. 1. Schematic of the computational domains with conjugate heat transfer solid domain).

*3.2. Effect of the homogeneous wettability of superheat wall on boiling curve*

In this section, the boiling process occurred in the homogenous wall is investigated, and the effects of the hydrophilic and hydrophobic walls on boiling curves are discussed in detail. It is worth mentioning that the adjustment of the different wettability is realized by a specific value of $G_w$ in Eq. (12) based on pseudopotential fluid-solid interaction force [44, 50].

3.2.1. Hydrophilic heating plate

Here, three kinds of hydrophilic surfaces are taken into consideration in current boiling processes, corresponding to $\theta=50°$, $\theta=60°$ and $\theta=70°$. The snapshots of boiling process at $t^*=57.63$ and $67.80$ with the contact angle of $\theta=50°$ are given by Figs. 2 and 3, respectively, for the wall superheat associated with $Ja=0.328, 0.373, 0.419$ and $0.487$. It should be noted that the high wall superheat is added after $t^*=5.08$. As can be seen in Figs. 2 and 3, the boiling processes occurred at $Ja=0.328$ and $Ja=0.373$ can be confirmed as the nucleate boiling regime due to the isolated small bubbles generated on the heating surface. In the meanwhile, as shown in Fig. 2(d), large part of heating surface covered the thin vapor film is the transition boiling due to the part of dryout. Therefore, it could be clearly concluded that, in this case, the boiling process has changed from the nucleate boiling to the transition boiling with the increase of wall superheat from $Ja=0.328$ to $Ja=0.487$.

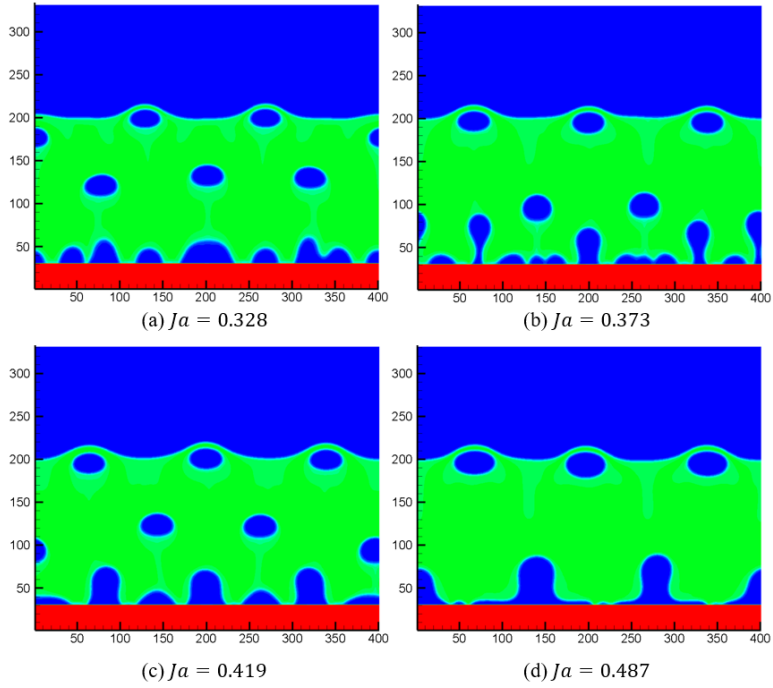

Fig. 2. Transient boiling characteristics at $t^*=57.63$ with different wall degrees of superheat ((a) $Ja=0.328$, (b) $Ja=0.373$, (c) $Ja=0.419$ and (d) $Ja=0.487$) under the hydrophilic surface of $\theta=50°$.

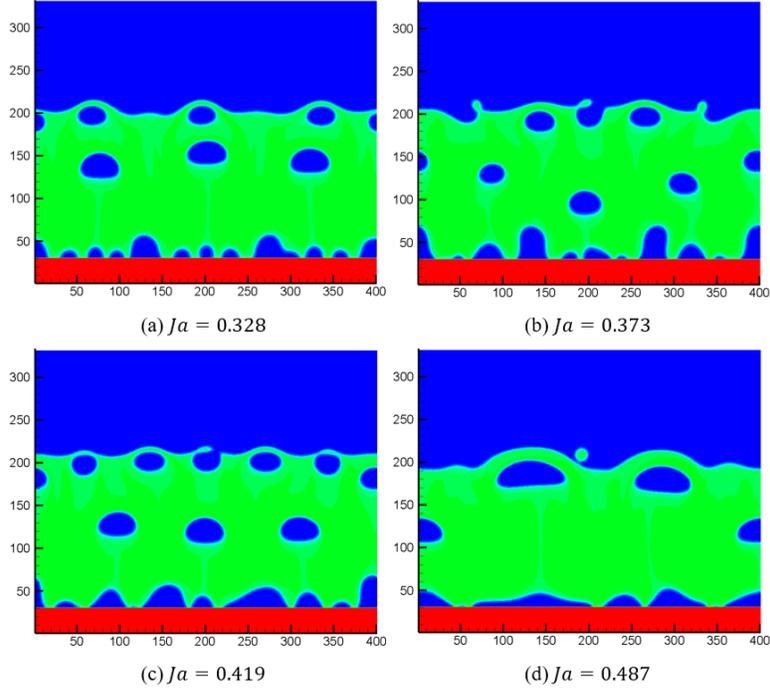

Fig. 3. Transient boiling characteristics at the same time of $t^*=67.80$ with different wall degrees of overheat ((a) $Ja=0.328$, (b) $Ja=0.373$, (c) $Ja=0.419$ and (d) $Ja=0.487$) under the surface's wettability of $\theta=50°$

The boiling characteristics occurred in the wall overheat from $Ja=0.328$ to $Ja=0.487$ with the wettability $\theta=60°$ are given by Fig. 4 at the same dimensionless time with Fig. 3. From this figure, one can observe that, the boiling process at $Ja=0.328$ and $Ja=0.373$ are also the nucleate boiling regime, while at $Ja=0.419$ and $Ja=0.487$, the boiling process can be clearly confirmed as the transition boiling and film boiling patterns, respectively. It means that as the wall wettability decreases (the contact angle increase), the boiling process has moved to film boiling with a low degree of superheat when comparing the Fig. 3 (d) and Fig. 4(d) at $Ja$=0.487. The boiling patterns in Fig. 4 can be further defined by the temporal variations of dimensionless heat flux versus dimensionless time for the four cases as demonstrated by Fig. 5. That is the transition boiling regime has a high fluctuate heat flux, whereas there is a quite stable transient heat flux associated with film boiling, which had also been confirmed by former literature [4, 29].

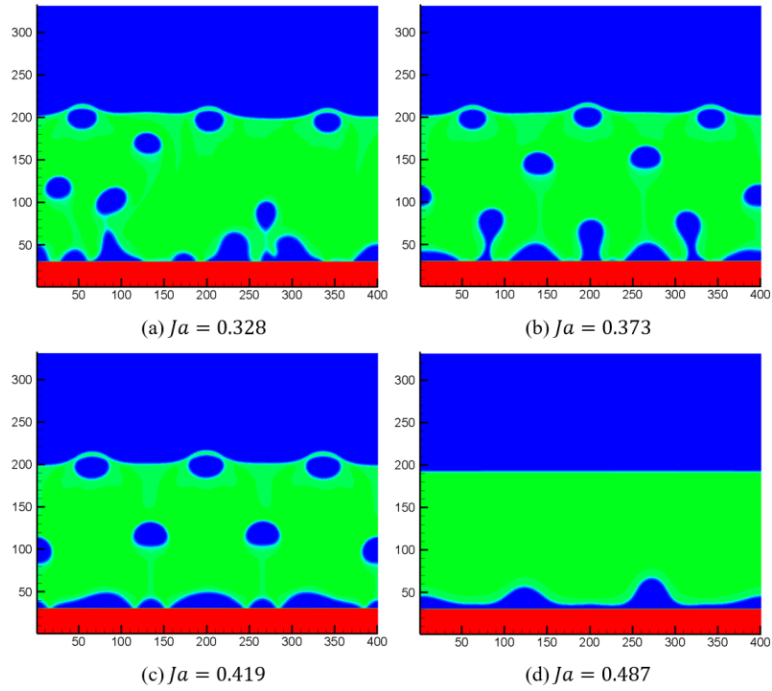

Fig. 4. Snapshots of boiling processes at the same time of $t^*$=67.80 with different wall superheats ((a) $Ja$=0.328, (b) $Ja$=0.373, (c) $Ja$=0.419 and (d) $Ja$=0.487) under the surface wettability of $\theta$ =60°.

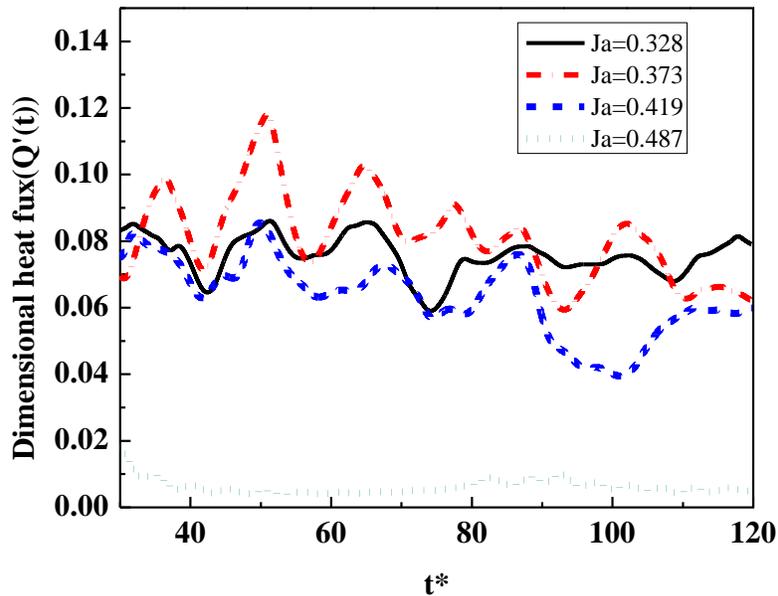

Fig. 5. Temporal variations of dimensionless heat flux versus dimensionless time for different wall overheats during the boiling processes under the surface's wettability of $\theta=60°$.

We further reduce the heating wall wettability to $\theta=70°$. The boiling processes for the cases Ja=0.328, 0.373, 0.419 and 0.487 at the same time with Fig. 5 are shown in Fig. 6. From Fig. 5, one can find that, the transition boiling regime has appeared at Ja=0.373 due to part of heating surface coved by thin vapor film, and film boiling has occurred at Ja=0.419. It indicates that the boiling processes have a shorter translation boiling regime compared to Figs. 3 and 5 when further decreasing the wettability of heating surface. Such a feature also demonstrates the last conclusion that with the augment of the contact angle, the boiling process can quickly move into the film boiling process with a lower degree of wall overheat. This is attributed to that decreasing the wall wettability would be beneficial to the vapor agglomeration at the heating surface, and this is also consistent with result concluded by Li et al.'s study [29].

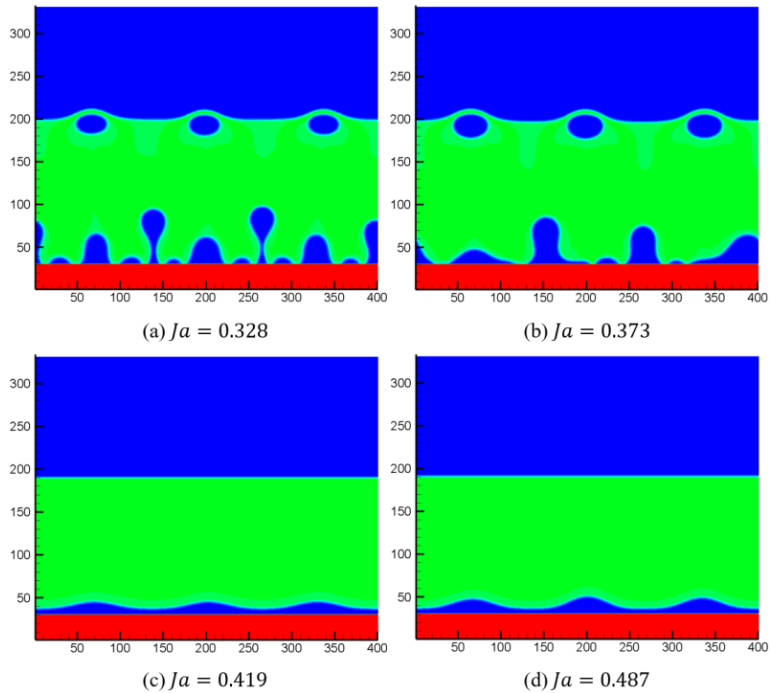

Fig. 6. Transient boiling characteristics at time of $t^*$=67.80 for the cases (a) Ja=0.328, (b) Ja=0.373, (c) Ja=0.419 and (d) Ja=0.487 with the wettability of $\theta=70°$.

In order to clearly picture the influence of wall wettability on the boiling process, the entire boiling curves for the cases of $\theta=50°$, 60° and 70° are plotted in Fig. 7 with the wall superheat from Ja=0.079 to Ja=0.555. As shown in Fig. 7, three simulated boiling curves for the three cases exhibit a good agreement with the classical boiling curve conducted by Nukiyama [3]. Note that, the dot-dashed lines B, C, D and E also correspond to the same wall superheat as Figs. 3 ,4 and 6, respectively. Therefore, the boiling regimes in Figs. 3, 4 and 6 can be further confirmed by Fig. 7. From Fig. 7, some critical results can be found. First, we can find that the critical heat flux (CHF) augments with the increase of wall wettability, while the requirement of the corresponding wall overheat increases accordingly. These results had also been demonstrated by former literature using the theoretical analysis [63], experimental studies [64, 65] and numerical investigation [29]. Besides, the above-mentioned conclusion that decreasing the wall wettability would result in a shorter transition boiling regime, can be clearly observed form Fig. 7. It indicates that the boiling characteristic can quickly move into the film boiling with a lower degree of wall overheat. Such a feature had also been revealed by previous research via the numerical research [30] and experimental study [65].

Furthermore, it can be found that three boiling curves intersect each other neat at $Ja$=0.215. After the intersect point, the heat flux augments with the increase of the wall wettability. On the contrary, before the intersect point, the heat flux increases as the wall wettability decreases. Such a feature is attributed to the different effect of contact angle on the ONB and CHF. It is also consistent with former study using experiment [11, 66] and numerical research [29]. In order to demonstrate the difference in the ONB, Fig. 8 provides the time evolutions of density contours under the wall superheat $Ja$=0.173 corresponding to line A. Note that, the left column is the case $\theta$=50°, while the right column is the case $\theta$=70°. As shown in Fig. 8, it can be observed that, the nucleate boiling has appeared in the case $\theta$=70°, whereas there is no phase-change formation in the case $\theta$=50°. Thereby, this phenomenon proves the above-mentioned conclusion that it is beneficial to produce ONB as wall wettability decreases. Moreover, from the entire boiling curves in Fig. 7, it is found the three pool boiling curves almost collapse into one in the film boiling patterns, and with the augment of wall wettability, the requirement of wall superheat to generate film boiling increase. This phenomenon is due to the fact that the transition boiling pattern of the pool boiling curves moves to the rightward shift as the wall wettability increases. It could be concluded that the wall wettability has a slight effect on the film boiling when the wall superheat is too high.

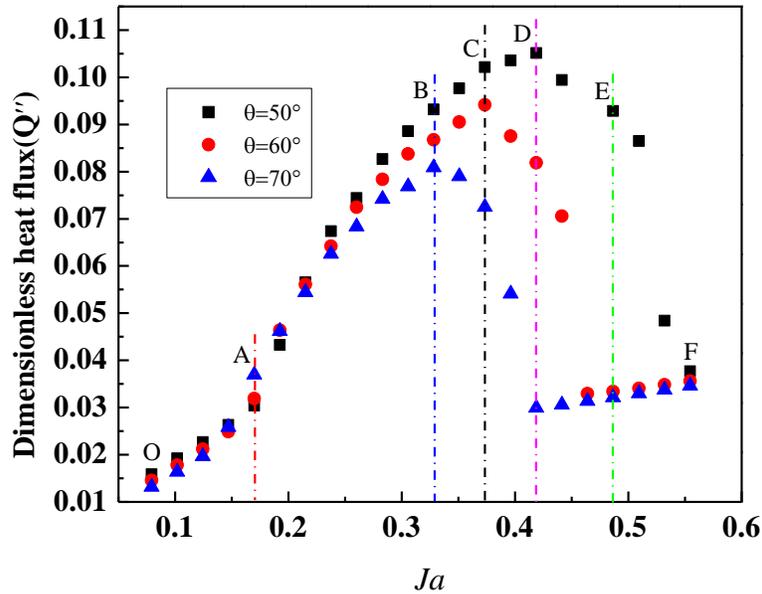

Fig. 7. Effects of surface wettability on the boiling curves under the hydrophilic heating heater ($\lambda_S/\lambda_L \approx 75$, $\lambda_L/\lambda_V \approx 17$) with the contact angle of cases (a) $\theta$=50°, (b) $\theta$=60° and (c) $\theta$=70°.

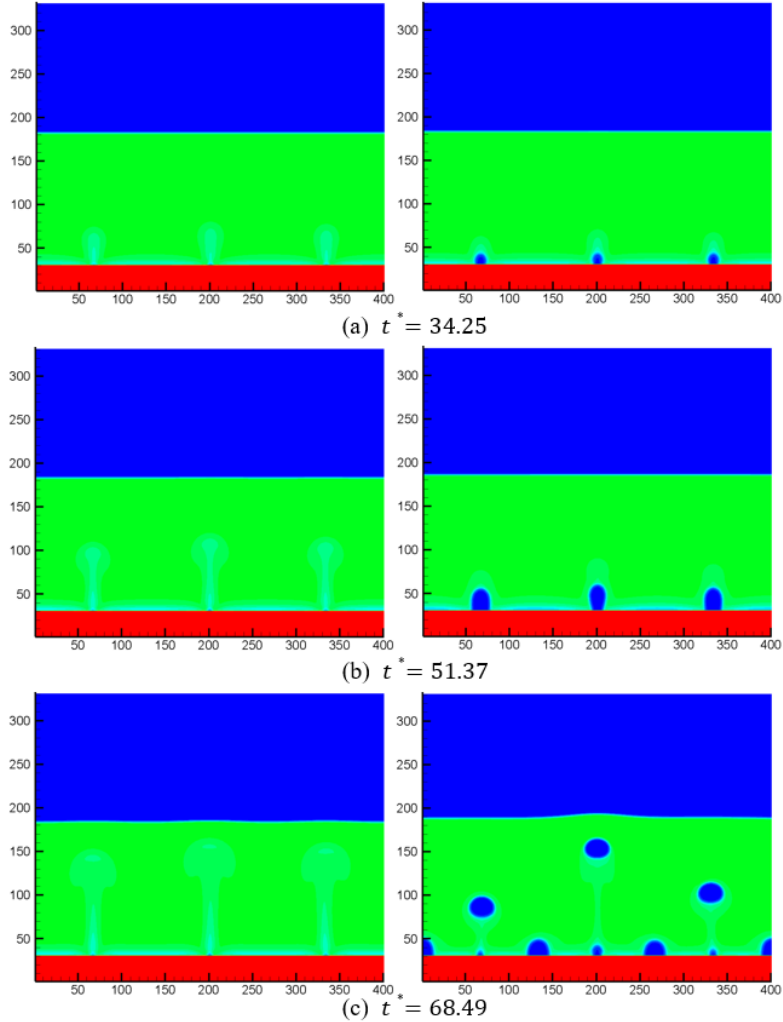

Fig. 8. Time evolution of boiling processes on the hydrophilic heating surface under the same wall superheat ($Ja$=0.170) with different wettability (the contact angle of left case is $\theta$ =50° and the right case is $\theta$ =70°)

3.2.2. Hydrophobic heater

In this part, the pool boiling processes occurred in hydrophobic heating surface are investigated. The snapshots of boiling processes with contact angle $\theta$ =95°, and 100° at $t^*$=74.58 are illustrated in Figs. 9 and 10, respectively, corresponding to the four cases of $Ja$=0.170, 0.215, 0.238 and 0.260. As shown in Fig. 9, the boiling process change from nucleate boiling pattern to translation boiling regime with the increment of the wall superheat. However, regarding the pool boiling occurred in heating plate with $\theta$ = 100°, the film boiling regime has appeared early instead of the transition boiling regime when comparing the boiling process in Figs. 9(d) and 10(d), at the same wall superheat $Ja$=0.260. These results imply that the boiling process can immediately move into the film boiling regime as the wall wettability of the hydrophobic heating surface slightly decreases.

Fig. 11 further provides the density contours at $t^*$=74.58 with wall wettability $\theta$ = 105°, for the cases $Ja$=0.170, 0.215, 0.226 and 0.238. As show in Fig. 11, the film boiling regime has taken place much earlier than above two heating situations at the same wall superheat $Ja$=0.238. Such a feature is caused by that the liquid is quite easier to detach from the heating surface due to the poor wall wettability, leading to the film boiling generated with a low wall superheat.

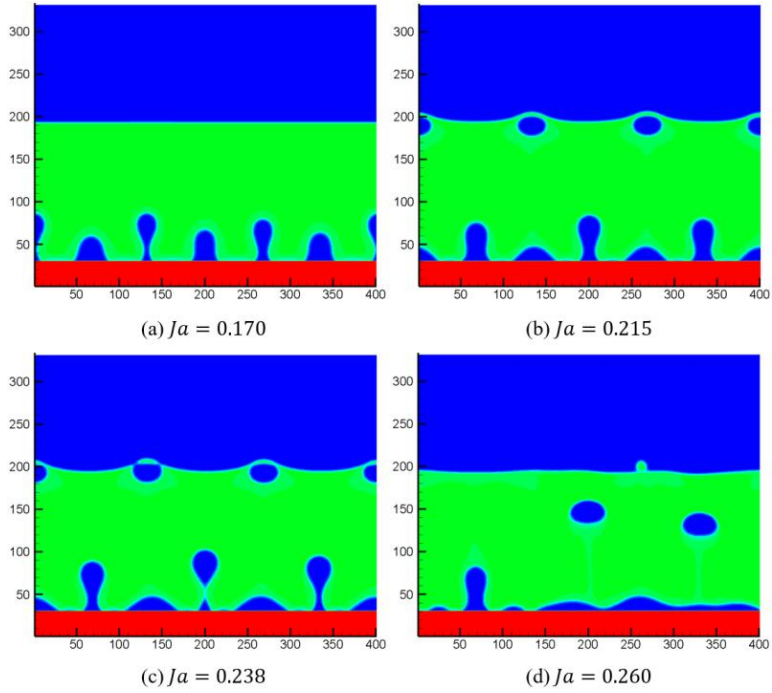

Fig. 9. Transient boiling characteristics at the same time of $t^*=74.58$ for the cases (a) $Ja=0.170$, (b) $Ja=0.215$, (c) $Ja=0.238$ and (d) $Ja=0.260$ with a slightly hydrophobic surface ($\theta =95°$).

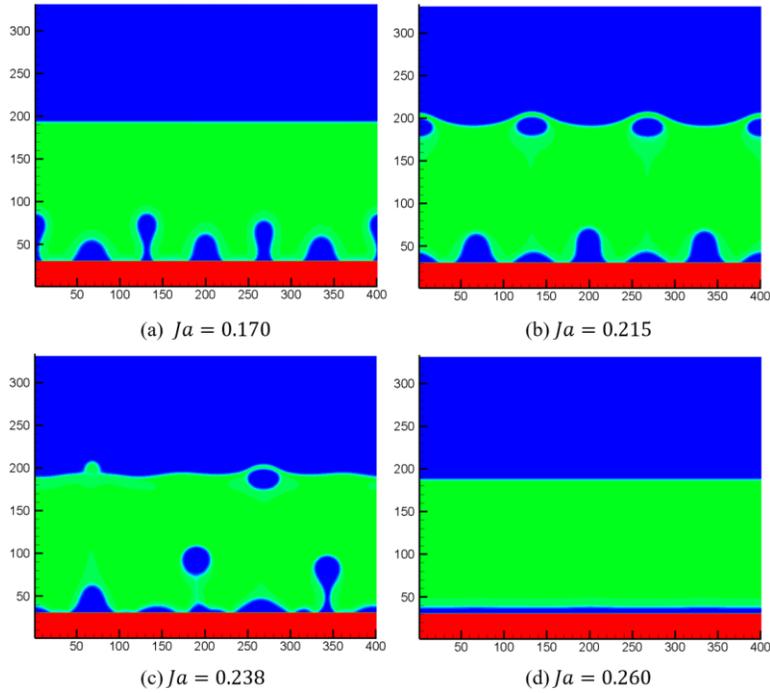

Fig. 10. Snapshots of boiling characteristics at the same time of $t^*=74.58$ for different wall superheats (a) $Ja=0.170$, (b) $Ja=0.215$, (c) $Ja=0.238$ and (d) $Ja=0.260$ with a hydrophobic surface ($\theta =100°$).

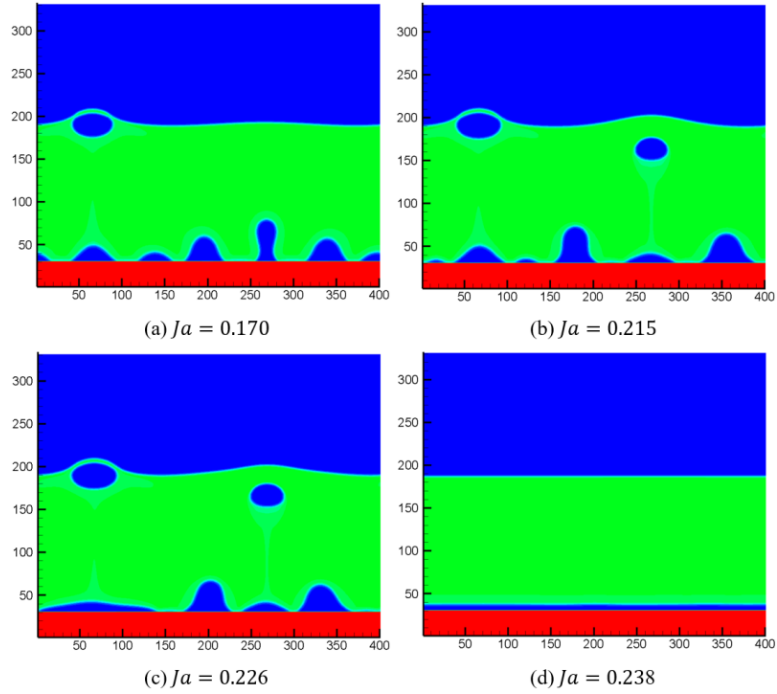

Fig. 11. Snapshots of boiling characteristics at the same time of $t^*=74.58$ for different wall superheats (a) $Ja=0.170$, (b) $Ja=0.215$, (c) $Ja=0.226$ and (d) $Ja=0.238$ with a hydrophobic heating surface ($\theta=105°$).

Similarly, for comparison, the three boiling curves obtained by $\theta=95°$, $100°$ and $105°$ are plotted in Fig. 12. It is worth mentioning that the dot-dashed lines A, B, C and D also correspond to the same wall superheat as Figs. 9 and 10, respectively. As a result, the boiling regimes in Figs. 9 and 10 can be further confirmed by Fig. 12. Form Fig.12 some similar features with the Fig. 7 are also obtained. The simulated results show that, in Fig. 7, the CHF values at the cases $\theta=50°$, $60°$ and $70°$ are given by 0.105, 0.094, 0.081, respectively, while the CHF values for the cases $\theta=95°$, $100°$ and $105°$ are evaluated by 0.049, 0.046, 0.038, respectively. Therefore, due to the effects of wall wettability on the boiling characteristics, the CHF values in the cases of the hydrophobic heating surfaces are near the half time of the CHF values under the hydrophilic heating surface, which indicates that the property of wall wettability has a significant influence on the CHF values. This feature is also consistent with previous research from experimental study [20]. Furthermore, as shown in Fig. 12, we also observe that it is quite easier to generate the nucleate boiling and film boiling regime with a low wall superheat for the hydrophobic heating surface when comparing to the boiling curves occurred in the hydrophilic heating surface as plotted by Fig 7. In order to prove this conclusion, the compression of boiling processes under the hydrophilic ($\theta=70°$) and slightly hydrophobic ($\theta=95°$) heating surface at the same low wall superheat ($Ja=0.124$) are given by Fig. 13. From this figure, it can be clearly observed that the phase change has appeared for the hydrophobic heating surface, whereas in terms of the hydrophilic heater, the phase change does not generate, which indicate that the nucleate boiling occurred on the hydrophobic heater requires a much lower degree of wall overheat than the hydrophilic heater. This conclusion is also consistent with the previous literature carried out by experiment study [16, 18] and numerical simulation [25].

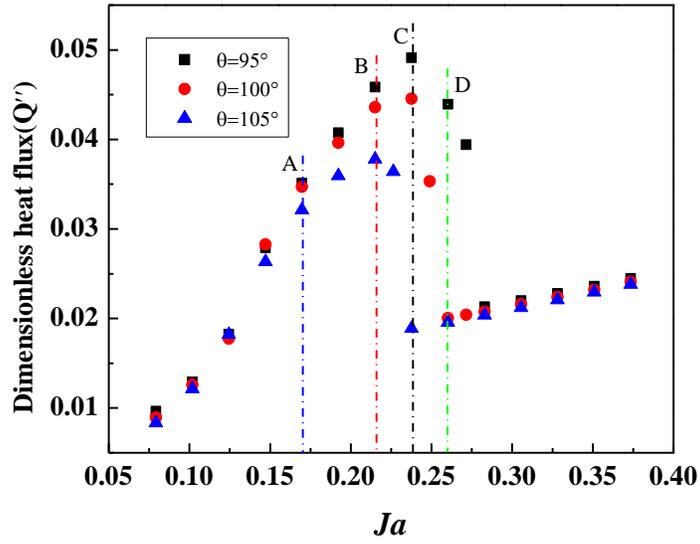

Fig. 12. Effects of surface wettability on the boiling curves under the hydrophobic heating plate ($\lambda_S/\lambda_L \approx 75$, $\lambda_L/\lambda_V \approx 17$) with the contact angle of cases (a) $\theta = 95°$, (b) $\theta = 100°$ and (c) $\theta = 105°$.

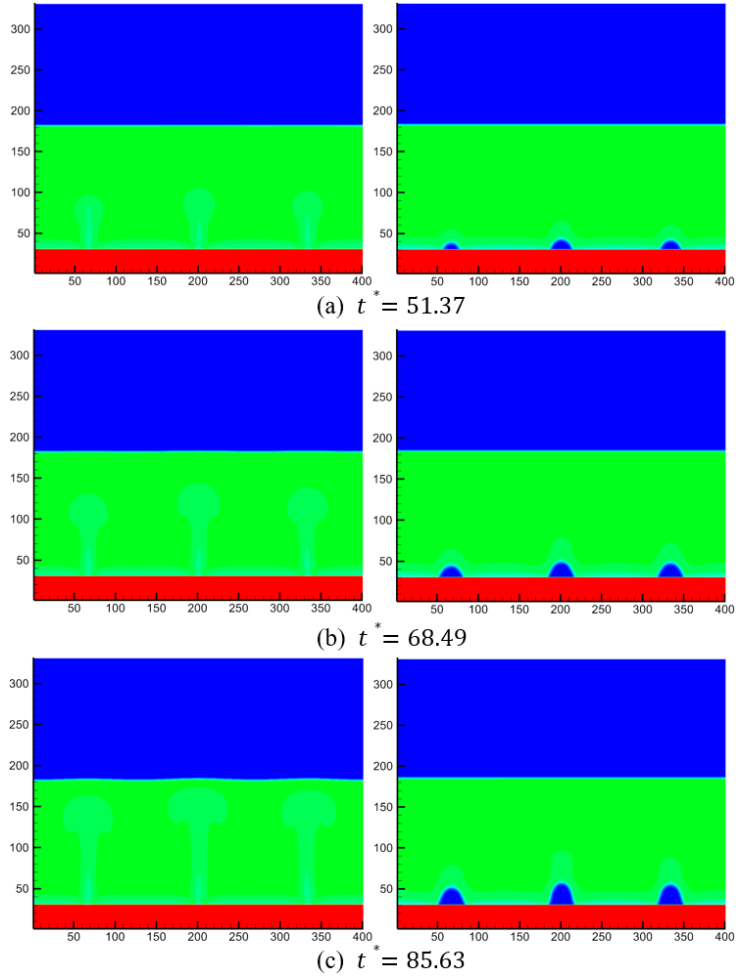

(a) $t^* = 51.37$

(b) $t^* = 68.49$

(c) $t^* = 85.63$

Fig. 13. Time evolution of boiling processes on the hydrophilic and hydrophobic heating plate (the contact angle of left case is $\theta=70°$ and the right case is $\theta=95°$) under the same wall superheat (*Ja*=0.124)

*3.3. Effect of the heterogeneous and hydrophilic wettability of superheat wall on boiling curve*

The above two sections are for the boiling process on the homogenous heater, therefore to further investigate the effect of mixed wettability on boiling process, in this subsection, the influence of the heating plate with heterogeneous and hydrophilic wettability on boiling curves is studied. The schematic structure with mixed wettability is illustrated in Fig. 14. As shown in this figure, the structure of the heater is set up with mixed wettability corresponding to wettability A $\theta_{h1}$ (denoted by blue color) and wettability B $\theta_{h2}$ (denoted by red color). Note that both two wall wettability are hydrophilic surface in this section. The boiling process on hydrophilic-hydrophobic mixed wettability will be discussed in next section. The width between two wall wettability is set to be L=40 l.u., while the width of the sub-structure wettability in the two ends of the computational domain is equal to L=20 l.u. in order to carry out the periodic boundary as shown in Fig. 14. Subsequently, two kinds of mixed wettability case a: $\theta_{h1}=50°$, $\theta_{h2}=60°$ and case b: $\theta_{h1}=50°$, $\theta_{h2}=80°$ are taken into consideration in current section. Two boiling curves with the wall superheat form *Ja*=0.079 to *Ja*=0.53. are numerically obtained and compared in Fig. 15. As shown in this figure, the maximum value of CHF increases with the augment of the length between two wall wettability. Meanwhile, the transition boiling regime moves to the leftward shift as the length between two wall wettability increases, resulting in the occurrence of film boiling regime with a lower wall overheat as displayed in line A shown in Fig. 15. Furthermore, increasing the length between two wall wettability would generate a high heat flux in the initial stage of nucleate boiling regime.

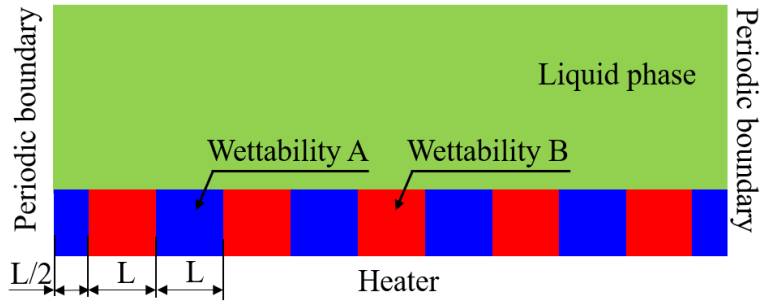

Fig .14. Schematic illustration of the heating surface with mixed wettability.

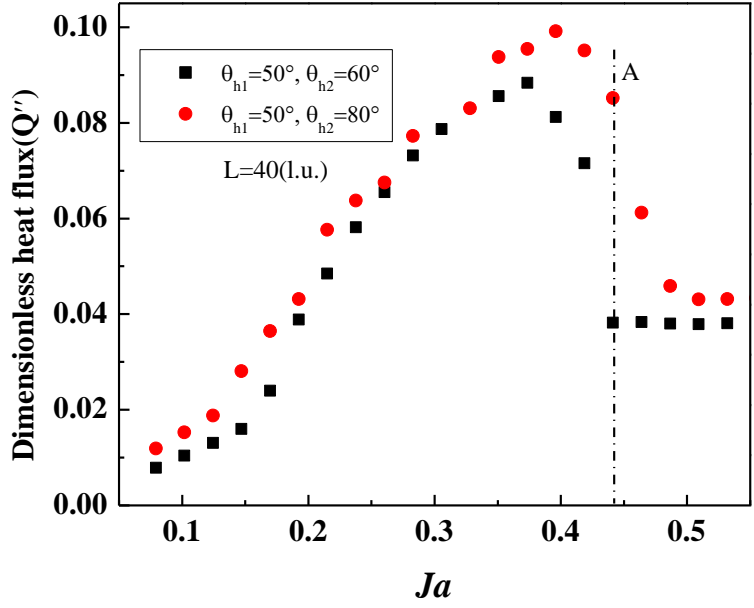

Fig. 15. Comparison of boiling curves on the heterogeneous heater ( $\lambda_s/\lambda_L \approx 75$ , $\lambda_L/\lambda_V \approx 17$ ) with different wettability for the case (a) $\theta_{h1}=50°, \theta_{h2}=60°$ and case (b) $\theta_{h1}=50°, \theta_{h2}=60°$ .

*3.4. Effect of the hydrophilic-hydrophobic mixed wettability of superheat wall on boiling curve*

In this part, we turn our attention to the boiling process under the hydrophilic-hydrophobic mixed wettability. Effects of the wall wettability of hydrophobic surface and the width between the mixed wettability surface on the entire boiling curves are numerically investigated. The schematic structure of the mixed wettability heating surface is also kept same as the Fig. 14.

3.4.1. Effect of the heating surface wettability

First, Effects of the wall wettability of the hydrophobic surface on boiling curves are studied. Three kinds of hydrophilic-hydrophobic mixed wettability are chosen as follow: cases (a) $\theta_h=60°, \theta_b=95°$ , (b) $\theta_h=60°, \theta_b=105°$ and (c) $\theta_h=60°, \theta_b=105°$ , which means that we just change the wettability of hydrophobic surface and fixing the wettability of hydrophilic surface. Note that the subscripts *h* and *b* denote the hydrophilic and hydrophobic surface, respectively. The width between two wall wettability is also chosen as L=40 l.u..

Three boiling curves under the controlled wall temperature are given by Fig. 16 for the hydrophilic-hydrophobic mixed wettability heater. It can be clearly seen from Fig. 16 that the heat flux increases with the decrease of the wall wettability of the hydrophobic surface at the same wall superheat throughout the entire boiling curves. In other words, the shapes of boiling curves move to the upward shift as the wall wettability of hydrophobic surface decreases from the nuclear boiling regime to the fully developed film boiling regime. Actually, such a feature is due to the fact that for the hydrophilic-hydrophobic mixed heating surface, increasing the contact angle of the hydrophobic surface will be contributed to the agglomeration of vapor at the heating surface and the departure of bubbles during the boiling process. By further insight into Figs.7 and 16 that, the obtained CHF value given by $\theta=60°$ in Fig. 7 under the homogenous heater is 0.094, while for hydrophilic-hydrophobic mixed wetting surface, the maximum value of CHF can reach to 0.117 when $\theta_h = 60°, \theta_b = 115°$. There is around 24.5% enhancement for the CHF occurred in the hydrophilic-hydrophobic mixed wall compared to the hydrophilic and homogenous wettability. Thereby, it could be concluded that the enhancement of boiling heat transfer for the hydrophilic-hydrophobic mixed surface is quantitively verified in current hybrid phase change LB model.

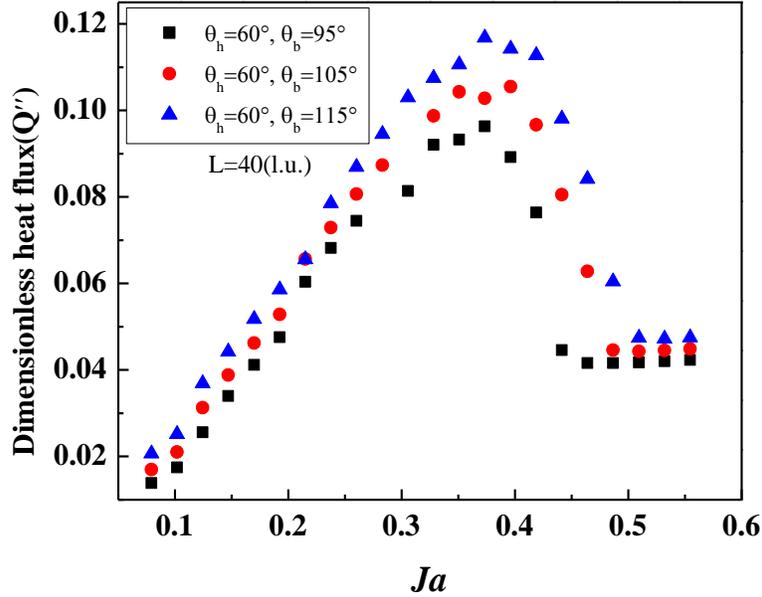

Fig. 16. Comparison of pool boiling curves on mixed hydrophilic-hydrophilic heater ($\lambda_S / \lambda_L \approx 75$, $\lambda_L / \lambda_V \approx 17$) for the cases (a) $\theta_h = 60°, \theta_b = 95°$, (b) $\theta_h = 60°, \theta_b = 105°$ and (c) $\theta_h = 60°, \theta_b = 115°$.

3.4.2. Effect of the width between two mixed wall wettability

Furthermore, the effects of the width between two wall wettability are also studied in this section. Two boiling curves under the heating surface having the hydrophilic-hydrophobic mixed wettability surface $\theta_h = 60°, \theta_b = 105°$ are numerically obtained and compared in Fig. 17, for the cases (a): L=20 l.u. and (b): L=40 l.u.. As shown in two boiling curves in Fig. 17, it can be clearly seen that three is a significant difference on the boiling curves form the nucleate boiling regime to the transition boiling regime. As the width between two wall wettability decreases, the nucleate boiling regime moves into the leftward and upward shift of boiling curves for the case (a). Additionally, decreasing the width between two wall wettability will leads to a leftward and upward shift of boiling curve for the case (a). This phenomenon is because that the different heat flux in the initial stage of nucleate boiling regime and the occurrence of CHF at difference wall superheat. First, to clearly picture the difference of initial nucleate boiling regime for two cases, Fig. 18 illustrates the transient boiling process for the two cases at the dimensionless time $t^*$=33.90 with wall superheat $Ja$=0.147, which also corresponds to line A in Fig. 17. From Fig. 18, we can observe that there is more nucleation site density in the case (a) than that in the case (b) at the same dimensionless time. Therefore, it could be concluded there is much higher heat flux for the case (a) than case (b). Besides, it is also found that both two case almost have a same CHF, but regarding the case (b), the CHF would require a higher wall superheat than that in case

A. Therefore, decreasing the width between two mixed wettability heating surface would yield the occurrence of CHF with a lower degree of wall overheat. Moreover, it is much easier to move into the film boiling regime for case (a) than that in case (b) as demonstrated in line B. In summary, these main results lead to the above-mentioned feature for the two boiling curves.

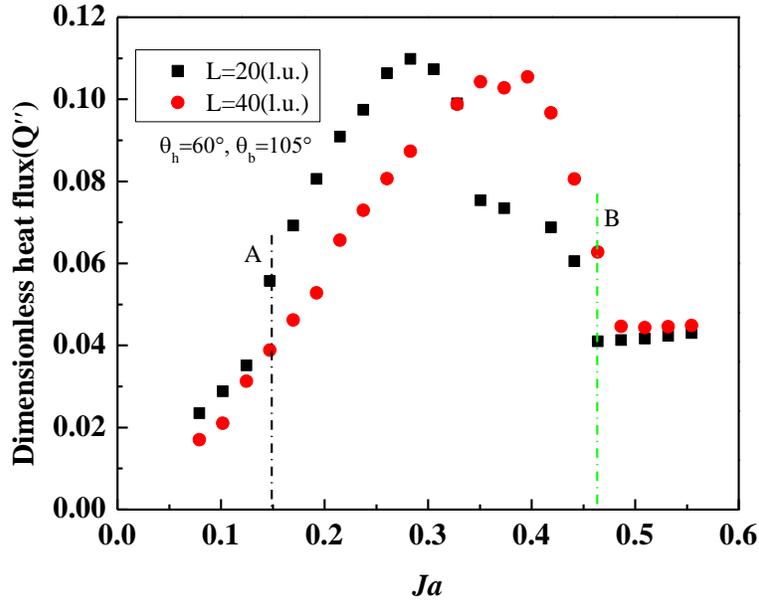

Fig. 17. Comparison of pool boiling curves on mixed hydrophilic-hydrophilic heater ($\lambda_S/\lambda_L \approx 75$, $\lambda_L/\lambda_V \approx 17$, $\theta_h = 60°, \theta_b = 105°$) with two structures for the cases (a) L=20 (l.u.) and (b) L=40 (l.u.).

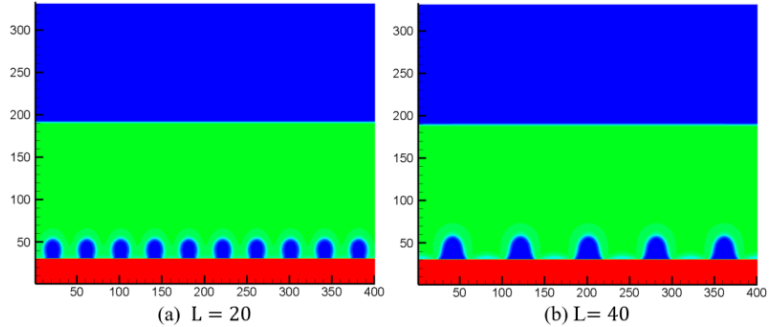

Fig. 18. Snapshots of boiling characteristics at the initial dimensionless time $t^*=33.90$ with a slightly low wall superheat $Ja$=0.147.

## 4. Conclusions

In this work, a hybrid pseudopotential phase change LB model is developed to direct numerical simulation of the saturated pool boiling processes under a smooth heating plate. The effects of wall wettability form the homogeneous surface, to heterogeneous surface to hydrophilic-hydrophobic mixed wettability wall on entire boiling curves are discussed in detail. The main conclusions of the current work are given below:

1. Regarding the boiling process occurred in the homogenous and hydrophilic heating plate, as the wall wettability augment, the CHF value increases, but it requires a high wall superheat accordingly. And decreasing the wall wettability results in the occurrence of ONB and film boiling patterns with a lower wall overheat. Furthermore, the pool boiling curves almost collapse into one in the film boiling patterns when the surface has a high wall overheat.
2. For the boiling process under the homogenous wall with hydrophobic wettability, there is much lower CHF value and a shorter transition boiling regime compared to the boiling process above the hydrophilic wall. However, it is quite easy to generate the nucleate boiling and film boiling with a lower wall superheat.
3. With respect to boiling process in the heterogeneous and hydrophilic heater, increasing the difference of the mixed wettability leads to the augment of CHF value and high heat flux in the initial nucleate boiling regime.
4. Comparison of boiling curves in terms of the hydrophilic-hydrophobic mixed wettability indicates that the shape of boiling curves moves into the upward shift with the increasing of the contact angle of the hydrophobic surface when the wettability of hydrophilic surface keeps the same.
5. Decreasing the width between the hydrophilic-hydrophobic mixed wall would result in a leftward and upward shift of the nucleate boiling pattern and a leftward and downward shift of the transition boiling regime.

## Acknowledgements

This work was supported by the National Natural Science Foundation of China (No.). Part of current work was done when the first author was working at West Virginia University supported by the China Scholarship Council (CSC, No.201806820023).